\newif\iftwocolumns
\twocolumnsfalse     

\iftwocolumns
   \documentclass[aip,jcp,reprint]{revtex4-1}
\else
  \documentclass[aip,jcp,preprint]{revtex4-1}
\fi

\usepackage{graphicx}
\usepackage{dcolumn}
\usepackage{bm}
\usepackage{amsmath, amssymb}

\def\const{\mathop {\rm const}\nolimits } 
\def\sn{\mathop {\rm sn}\nolimits } 
\def\K{\mbox{\rm K}} 
\def\F{\mbox{\rm F}} 
\def\E{\mbox{\rm E}} 

\newcommand{\Wr}{{\cal W}\hspace{-0.3mm}{\it r}} 
\newcommand{\Tw}{{\cal T}\hspace{-0.3mm}{\it w}} 
\newcommand{\Lk}{{\cal L}\hspace{-0.3mm}{\it k}} 

\begin{document}

\title{Condensation of circular DNA}

\author{E.L.Starostin}
\affiliation{ 
Department of Civil, Environmental \& Geomatic Engineering, University College London, Gower Street, London WC1E 6BT, UK}
\email{e.starostin@ucl.ac.uk}

\date{\today}

\begin{abstract}
A simple model of a circularly closed dsDNA in a poor solvent 
is considered
as an example of a semi-flexible polymer with self-attraction.
To find the ground states, the conformational energy is computed as a sum of
the bending and torsional elastic components and the effective self-attraction energy.
The model includes a relative orientation or sequence dependence of the effective attraction forces between
different pieces of the polymer chain.
Two series of conformations are analysed: a multicovered circle (a toroid) and a multifold two-headed racquet.
The results are presented as a diagram of state. It is suggested that the stability of particular conformations may be controlled
by proper adjustment of the primary structure. Application of the model to other semi-flexible polymers is considered.

\end{abstract}

\pacs{87.15.bk, 36.20.-r, 87.15.A-}

\maketitle

\section{\label{sec:intro} Introduction}


When a double-stranded DNA (dsDNA) finds itself in a solution with a condensing agent, it tends to take on a conformation that
would minimise its exposure to the environment and, correspondingly, maximise the sites of self-contact~\cite{Bloomfield96,Cherstvy11}.
Since dsDNA is a rather stiff molecule, variation of its shape implies a certain energy cost which is usually described
as the elastic energy of bending and twisting.
The aim of this paper is to derive simple estimates that help us to decide what conformation is preferable 
for a circularly closed dsDNA in terms of its total energy. 
The molecule is assumed to be unknotted and torsionally relaxed when in pure water.
We are interested in finding stable ground states and do not consider temperature effects.

\begin{figure}                
\iftwocolumns
\begin{center}
\resizebox{\columnwidth}{!}{\includegraphics{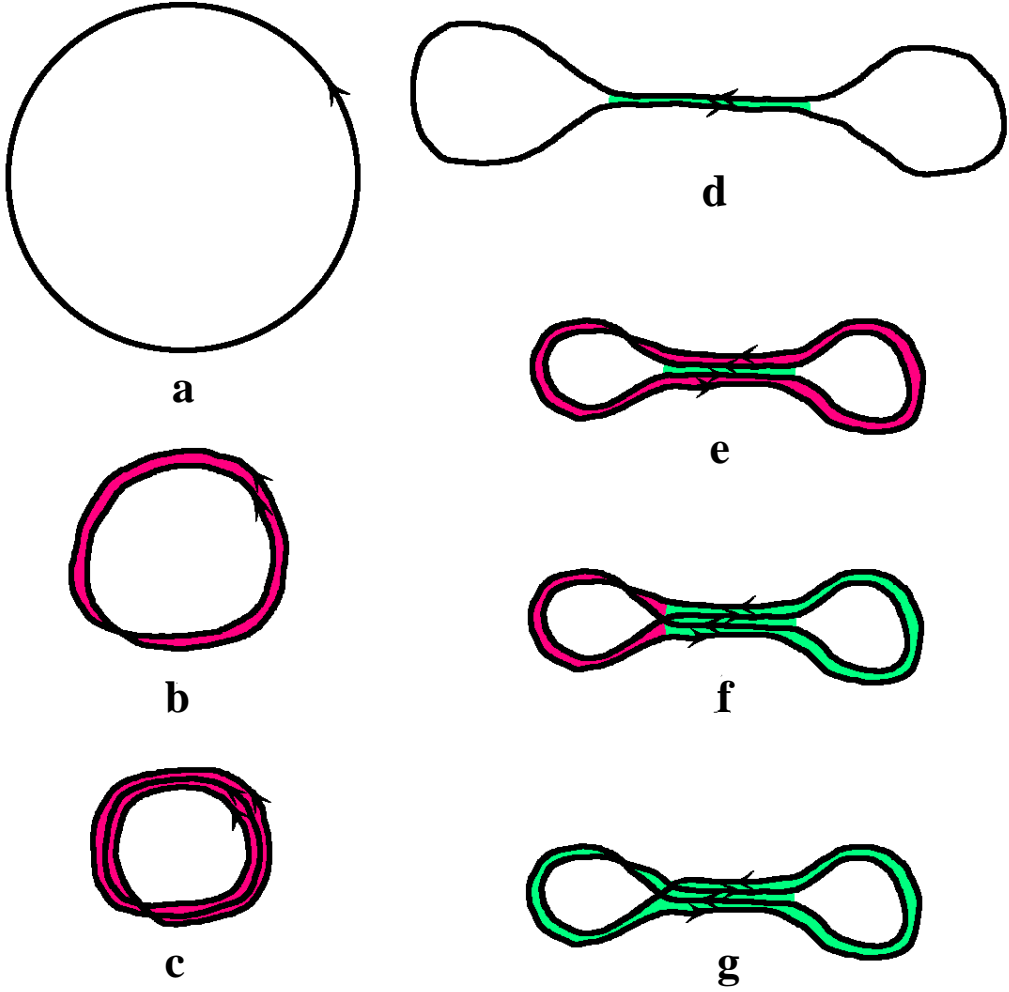}}
\end{center}
\else
\includegraphics[width=14cm]{scheme_4f.eps}
\fi
\caption{\label{fig:scheme} 
Two ways of condesation of a circular dsDNA: $N$-covered circles (toroids) (a: an initial conformation with $N=1$, b: $N=2$, c: $N=3$) and $M$-folded two-headed racquets 
(d: $M=1$, e,f,g: $M=2$).
Red colour marks parallel self-alignment, green antiparallel.
The shapes (e), (f) and (g) differ with the writhing number, e: $\Wr =\pm 1$, f: $\Wr=\pm 1 \pm 1$, g: $\Wr=\pm 1$ (unknotted).}
\end{figure}

If we assign orientation to the centreline of the dsDNA, then two ways of condensation are logically possible: 
the remote intervals of the dsDNA chain can align next to each other such that their orientation is either the same or opposite.
The first case is best achieved in a double- (or multiple-)covered circle, while the second realises 
as a straight line interval with two loops at the ends (Fig.~\ref{fig:scheme}).
Thus we come to familiar toroidal and racquet-like shapes~\cite{Schnurr00,Schnurr02}.
Here we consider the series of the simplest possible conformations: a toroid and a two-headed racquet.
The latter does not necessarily imply a significant writhe, while the first has its writhing number growing linearly with
its number of coils~\cite{Fuller71}.
Thus, the writhing number of a toroid with $N$ coils equals $\pm(N-1)$, if we neglect the thickness of the polymer,
while a multiple-covered two-headed racquet may be made
to have its writhing number vanish.
We assume that the linking number is conserved, therefore, the non-zero writhe is compensated by twist 
which should be accounted for in the calculation of the elastic energy.
Recall that, to avoid accumulation of twist in long ropes, sailors store them in a figure-8 fashion, a similar trick is used when mountaineers ``butterfly'' the coils of their climbing ropes. 
These coilings are analogous to the two-headed racquets.

Change of the relative orientation of contacting pieces of dsDNA is only one, most pronounced, feature
of two types of condensate. Actually, the coupling of the particular sites that interact with each other differ as well.
On the other hand, the interaction potentials of two homologous and nonhomologous DNA duplexes 
are found to be significantly different~\cite{Kornyshev07}.
This suggests that the monomer sequence may be intentionally tuned to amplify the effective attraction forces 
for one specific conformation. 
Toroids and racquets may be considered as analogs of the secondary structures of single-stranded nucleic acids
where dsDNA plays part of a single stranded polymer and the hydrogen bonds between complementary nucleotides are replaced with
the electrostatic attraction. The secondary structure of RNA or ssDNA may be controlled by varying their primary structure.
Similarly, the primary structure may affect the relative stability of toroids and racquets as the dsDNA condensates. 
Note that the sequence-dependent DNA condensation was studied on double-stranded poly(dG-dC)$\cdot$(dG-dC) (GC-DNA) 
and ds poly(dA-dT)$\cdot$(dA-dT) (AT-DNA)~\cite{Sitko03}. The effects due to the relative orientation of the contacting
pieces or to their specific sequence contents do not exist for such homogeneous primary structures.

As known from earlier work (both theoretical and experimental) for open polymer chains~\cite{Schnurr00,Schnurr02,Sakaue02,Sitko03,Montesi04}, 
the racquets can only be so-called metastable structures which means that they always lose to
toroids though their conformational energy may not differ much from that of toroids.
Numerical simulation suggests that interaction with adsorbing surface tend to stabilise metastable racquet-like structures~\cite{Barsegov05}.
Experiments with condensation of circular plasmids (closed un/nicked and open)
also show intermediate racquet-like shapes which seem to occur more often for unnicked DNA~\cite{Boettcher98}.
What if the racquets be given the upper hand by building a DNA with specially tuned mirrored homologous subsequences
in order to increase their interaction potential? The competing toroids must not feel it which
could cause loss of their role as most stable structures.
The model considered in this paper explores such a possibility. 
To simplify analysis it will be assumed
that all the contacting sites in racquets have uniform interaction potential which can differ from that of toroids where
the interaction forces are also assumed to be uniform.
It was recently demonstrated by numerical simulation that
presence of a direction-dependent potential affects condensation of semi-flexible polymers~\cite{Englebienne12}.

\section{Conformational energies}

We use a model of the semi-flexible polymer chain that includes two parts of the conformational energy: the elastic energy and the self-interaction energy~\cite{Schnurr02}.
The balance between these two components determines the geometry of the ground-state configuration.

For a polymer molecule of length $L$, we describe the bending energy as that of a slender
elastic rod:
\begin{equation}
U_{bend}=\frac{B}{2}\int\limits_0^L \kappa^2(s)\,\mbox{d}s,
\label{eq:bend_en}
\end{equation}
where $B$ is the bending stiffness and $\kappa(s)$ is the curvature of the centreline.
Similarly, the torsional energy is computed as
\begin{equation}
U_{tors}=\frac{C}{2}\int\limits_0^L \omega^2(s)\,\mbox{d}s,
\label{eq:tors_en}
\end{equation}
where $C$ is the torsional stiffness and $\omega(s)$ is the twist rate of the polymer.
We do not account for dependence of the elastic moduli on the variations of the sequence of monomers.
Neither stretching of DNA is included in the model as of minor importance.

Each piece of the polymer chain may be in contact with up to six other pieces.
Condensation is possible when every such contact decreases the conformational energy.
In other words, the polymer prefers to contact with itself in order to screen its surface from exposition to the solvent.
To account for the self-attraction, we employ the concept of the coordination mumber $\alpha_N$.
The doubled coordination number $2\alpha_N$ counts the number of possible binding sites that are left non-occupied by $N$
parallel polymer strands which fill in the triangular lattice in the orthogonal cross section~\cite{Schnurr02}.
There exists an exact formula for the minimal coordination number~\cite{Starostin08c}
\[
\alpha_N=\left\lceil\sqrt{12 N - 3}\right\rceil,
\]
where $\lceil\cdot\rceil$ stands for the ceiling function: for $x \in \mathbb{R}$, $\lceil x\rceil$ is defined as the smallest integer greater than or equal to $x$.
In particular, $\alpha_1 = 3$, $\alpha_2 = 5$, $\alpha_3 = 6,$ and so on.
Let $\gamma$ be the energy gain of one contact per length. We shall assume that $\gamma > 0$.
Then for an $N$-fold parallel bundle of length $L_N$ the interaction energy is proportional to the difference
of the sum of the coordination numbers of $N$ separate chains and the coordination number of them packed together
\begin{equation}
U_{N}=-\gamma (\alpha_1 N - \alpha_N) L_N .
\label{eq:surf_en}
\end{equation}

In what follows, all the lengths and energies  will be presented in dimensionless form by normalising them
on the {\it condensation length} $L_c = \sqrt{B/\gamma}$ and
the {\it condensation energy} $U_c = \sqrt{B \gamma}$, respectively~\cite{Schnurr02}. We shall denote the normalised contour length
of the polymer chain by $\lambda = L/L_c$.
The analysis is carried out under a simplified assumption of the zero thickness of the polymer.

\subsection{Multi-covered circle (toroid)}

In this work we consider only circularly closed polymers. The reference conformation is the single-covered circle which
minimises the elastic energy. DNA is torsionally relaxed so that the linking number 
of its strands $\Lk = \Lk_0 = \Tw_0$, where $\Tw_0$ corresponds to the intrinsic twist of the B-form.
The normalised bending energy is $u_{bend}=2\pi^2/\lambda$.
There is no self-interaction
and, correspondingly, Eq.~(\ref{eq:surf_en}) gives us zero for $N=1$.

The circularly closed DNA may form a $N$-covered circle or a toroid without being nicked.
The size of the toroid can only take discrete values contrary to the open case.
Let $\rho$ be the radius of the circle which is traced $N$ times by the DNA before it closes onto itself so that
$L = 2\pi\rho N$.
Then the curvature $\kappa = 1/\rho =\const$ and the bending energy Eq.~(\ref{eq:bend_en})
equals $U_{bend}=\frac{B}{2}\frac{1}{\rho^2} L = 2\pi^2 B \frac{N^2}{L}$.

Since the linking number is conserved, we may estimate the change of twist as
$\Tw-\Tw_0 = - \Wr$, where the writhing number of the $N$-covered circle can be estimated
as $\pm (N-1)$.
We further assume that the twist is uniformly distributed along the length of the polymer, i.e. $\omega(s) = \pm 2\pi (N-1) /L$.
Then the torsional energy can be found from Eq.~(\ref{eq:tors_en}):
$U_{tors}=\frac{C}{2} \omega^2(s) L = 2 \pi^2 C \frac{(N-1)^2}{L}$,
and after nomalisation we have 
$u_{tors} = 2 \pi^2 c  \frac{(N-1)^2}{\lambda}$, where we introduced
$c=C/B$. Note that in order to model a nicked DNA one can formally set $c=0$ to kill the torsional energy.

The fact that the polymer touches itself uniformly along all its length 
allows us to apply Eq.~(\ref{eq:surf_en}) for $L_N=L/N$. 
Then the normalised interaction energy is simply $u_{int}=(\alpha_N-N\alpha_1)\lambda/N$.
Summing up all three energies gives the total
\begin{align}
u_{t} &=u_{bend}+u_{tors}+u_{int}= \nonumber \\
&=\frac{2\pi^2}{\lambda} (N^2+c(N-1)^2) + \left(\frac{\alpha_N}{N}-3\right)\lambda .
\label{eq:en_tor}
\end{align}
The above expression coincides with Eq.~(4) in Ref.~\cite{Battle09} for $c=0$.

\subsection{Two-headed racquet}

To minimise the bending, the heads should be identical.
Then the elastic energy reduces to the doubled bending energy of a racquet head which is assumed to 
be a bundle of $M$ identical plane inflexional elasticae~\cite{Love27}. Let the contour length of the head be denoted by $\chi$ which is the only parameter
the normalised bending energy of a single component depends on:
$u_{head} = a/\chi$, where $a=4\xi^2(k_0)(2 k_0^2-1)$,
$\xi(k_0) = 2\K(k_0) - \F\left(\frac{1}{k_0\sqrt{2}},k_0\right)$~\cite{Schnurr02}. 
Here $\F$ denotes the elliptic integral of the first kind of modulus $k$
\[
\F(z,k)=\int\limits_0^z\frac{1}{\sqrt{1-t^2}\sqrt{1-k^2t^2}}\,\mbox{d}t,
\]
and $\K(k) \equiv \F(1,k)$ is the complete integral.
The elliptic modulus $k_0$ is found as a root of the equation $2\eta(k_0)=\xi(k_0)$, where
$\eta(k) = \E(\sn(\xi(k),k),k) = 2 \E(k) - \E\left(\frac{1}{k\sqrt{2}},k\right)$
and $\E$ is the elliptic integral of the second kind of modulus $k$
\[
\E(z,k)=\int\limits_0^z\frac{\sqrt{1-k^2t^2}}{\sqrt{1-t^2}}\,\mbox{d}t ,
\]
$\E(k) \equiv \E(1,k)$.
Numerical solution gives $k_0 \approx 0.8551$ and $a \approx 18.3331$.

To compute the interaction energy of the head, we apply Eq.~(\ref{eq:surf_en}) which becomes
$u_{M}=(\alpha_M - M \alpha_1) \chi$.
The straight handle is a $2M$-bundle of normalised length $\lambda/(2M)-\chi$ and Eq.~(\ref{eq:surf_en}) yields
$u_{2M} = (\alpha_{2M}-2M\alpha_1)(\lambda/(2M)-\chi)$.
The total energy is the sum $2M u_{head} + u_{M}+u_{2M}$.
Like for an open polymer, the size of the racquet head may take continuous values. 
For given length, the optimal length of the head equals $\chi=\sqrt{\frac{2aM}{\alpha_M-\alpha_{2M}+3M}}$ 
and finally we have $u=2\sqrt{2aM(\alpha_M-\alpha_{2M}+3M)} + (\alpha_{2M}/(2M)-3)\lambda$.
Note that the optimal racquet may only exist for polymers not shorter than $2M\chi$.
Its actual size is proportional, for given $M$, to the condensation length (contrary to discrete diameters of the toroids
made of a closed polymer).

To account for the difference in the effective attraction forces for parallel and antiparallel alignments 
we distinguish the interaction energy coefficient for the racquet $\gamma_{r}$.
It is convenient to introduce a new parameter $\nu$, $\nu^2 = \gamma_{r} / \gamma$ which measures
the orientation difference.
$\nu^2 >1$ corresponds to stronger attraction in the antiparallel configuration (as in the racquet handle). On a triangluar lattice the antiparallel arrangement is frustrated and in the ground state
one third of all neighbouring pairs have the same orientation, similarly to antiferromagnetic materials~\cite{Wannier50}.  When computing the value of $\nu$, this property
should be taken into consideration.

To keep the common normalisation, we rewrite the racquet energy as 
\begin{equation}
u_r=2\sqrt{2aM(\alpha_M-\alpha_{2M}+3M)} \nu + (\alpha_{2M}/(2M)-3)\lambda\nu^2 .
\label{eq:en_racq}
\end{equation}

\section{Comparison of conformations}

\begin{figure}                
\iftwocolumns
\begin{center}
\resizebox{\columnwidth}{!}{\includegraphics{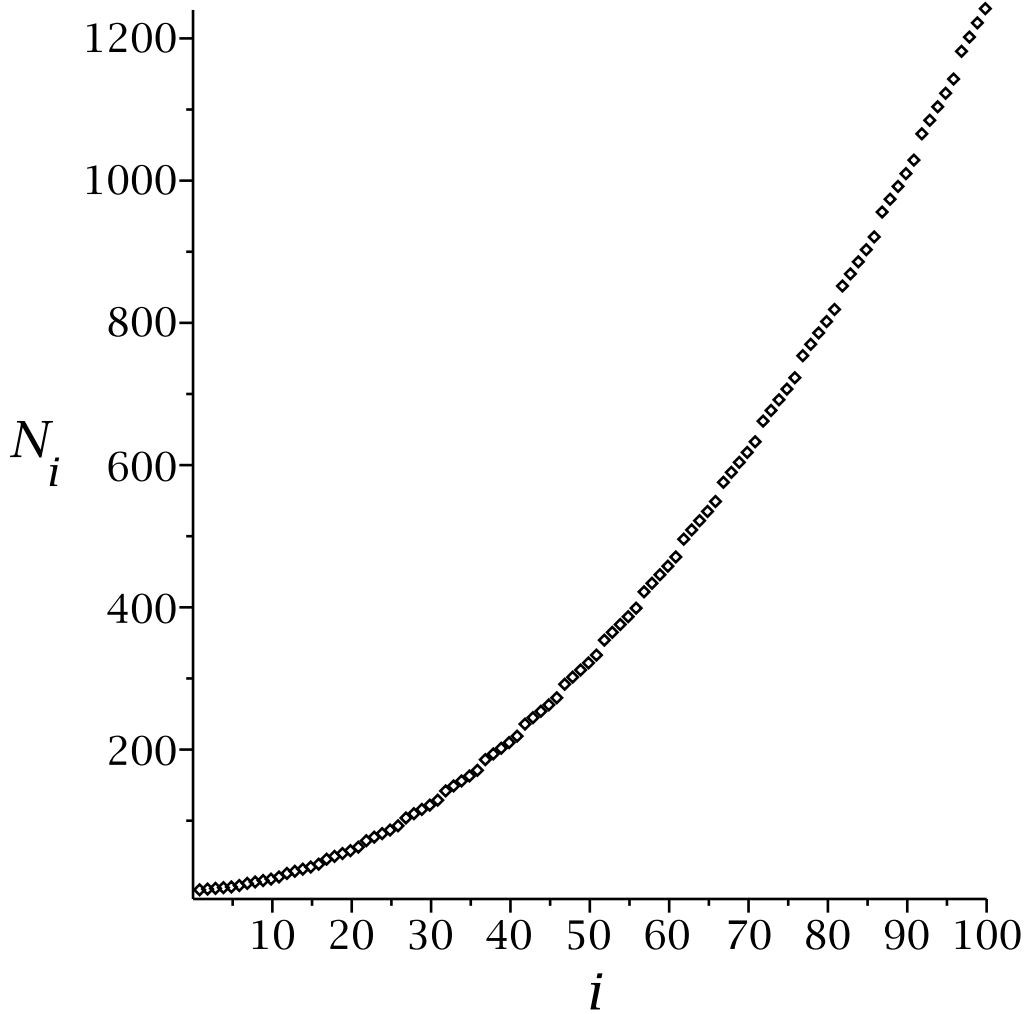}}
\end{center}
\else
\includegraphics[width=14cm]{toroids_N_i.eps}
\fi
\caption{\label{fig:N_i}
The sequence $\{N_i\}$ for the optimal $N_i$-covered circles (toroids). }
\end{figure}

\begin{figure}                
\iftwocolumns
\begin{center}
\resizebox{\columnwidth}{!}{\includegraphics{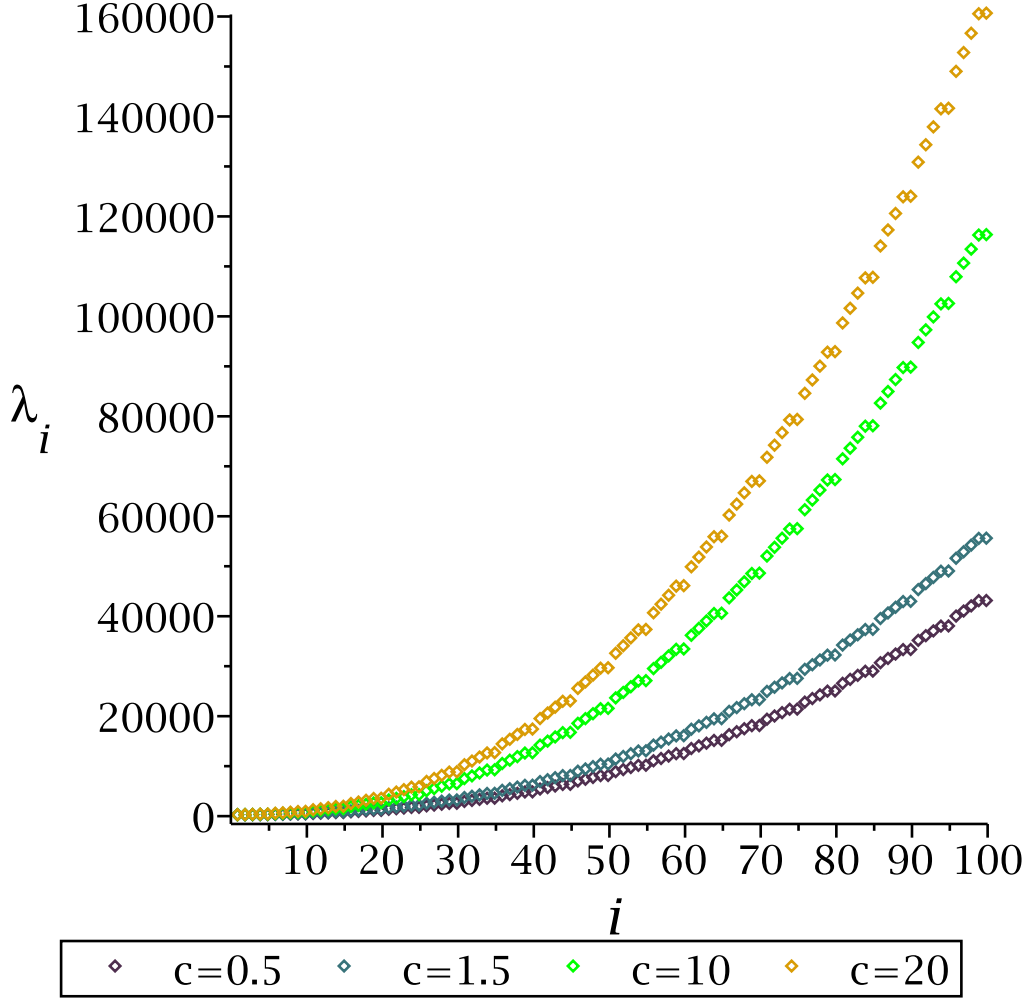}}
\end{center}
\else
\includegraphics[width=14cm]{toroids_lambda_i.eps}
\fi
\caption{\label{fig:lambda_i}
The sequence of critical lengths $\{\lambda_i\}$ for the optimal $N_i$-covered circles (toroids) for four different
values of the relative torsional stiffness $c$. }
\end{figure}

Now we are able to compare the energy of the conformations as functions of the polymer length and
the orientation parameter $\nu^2$.
We begin with $N$- and $P$-covered circles which are members of a sequence
of toroidal shapes.
Comparison of the conformational energies (Eq.~(\ref{eq:en_tor})) leads to the critical length
\begin{equation}
\lambda_{NP}^2 =2\pi^2 NP(N-P)\frac{(N+P)(c+1)-2c}{N\alpha_{P}-P\alpha_{N}} .
\label{eq:lambdaNP}
\end{equation}
By use of this expression we can build a sequence of optimal toroids $\{N_i\}$ with critical lengths $\{\lambda_i\}$. 
For a polymer of length $\lambda$, the $N_i$-circle is optimal among all toroidal shapes
if $\lambda_{i-1} < \lambda < \lambda_{i}$.
We first set $N_0=0$, $\lambda_0=0$ and start with a circle with $N_1=1$.
Next, assuming that we know the optimal sequence up to $i=I$ (i.e. we know
$N_i$, $i=0,\ldots, I$, and $\lambda_{i}$, $i=0,\ldots, I-1$)
we compute $\lambda_I=\lambda_{N_I P^\star} =\min\limits_{P>N_I} \lambda_{N_I P}$,
$N_{I+1}=\max\limits_{\lambda_{N_I P^\star}=\lambda_I} P^\star$.
It is easy to check that $\partial_\lambda u_t(\lambda,P^\star) < \partial_\lambda u_t(\lambda,N_I)$
for $\lambda = \lambda_{I}$ and hence there exists $\epsilon >0$ such that
the inequality holds for $\lambda \in [\lambda_{I},\lambda_{I}+\epsilon)$.
Therefore the $N_{I+1}$-toroid is optimal for $\lambda > \lambda_{I}$ in some right neighbourhood of $\lambda_{I}$.
Now we can apply the same procedure to find $\lambda_{I+1}$, $N_{I+2}$ and so on. 
The first members of the optimal toroid sequence are (Fig.~\ref{fig:N_i}):
\begin{align}
S_t =\{&1,2,3,4,5, \nonumber \\
&7,10,12,14,16, \nonumber \\
&19,24,27,30,33, \nonumber \\
&37,44,48,52,56, \ldots\} .
\label{eq:seqtor}
\end{align}
The critical lengths are shown in Fig.~\ref{fig:lambda_i} for four different values
of the torsional to bending stiffness ratio $c$ (note that, as can be seen from
Eq.~(\ref{eq:lambdaNP}), the optimal sequence $S_t$
does not depend on $c$ for nonnegative values).

The sequence Eq.~(\ref{eq:seqtor}) may be described as a series of quintuples
\begin{align*}
S_t=\{ &3n(n+1)+1, \\
            & \{(3n+m+1)(n+1), m=1,2,3,4\}, \\
&n=0,1,2,\ldots\} .
\end{align*}
As proposed in Ref.~\cite{Battle09}, the sequence is made with the numbers that
satisfy $\alpha(N)=\alpha(N-1)=\alpha(N+1)-1$.
This characterisation must be corrected: 
numbers that can be expressed as $(n+1)(3n+1)$, $n=1,2,3,\ldots$,
must be removed from the optimal sequence.

\begin{figure}                
\iftwocolumns
\begin{center}
\resizebox{\columnwidth}{!}{\includegraphics{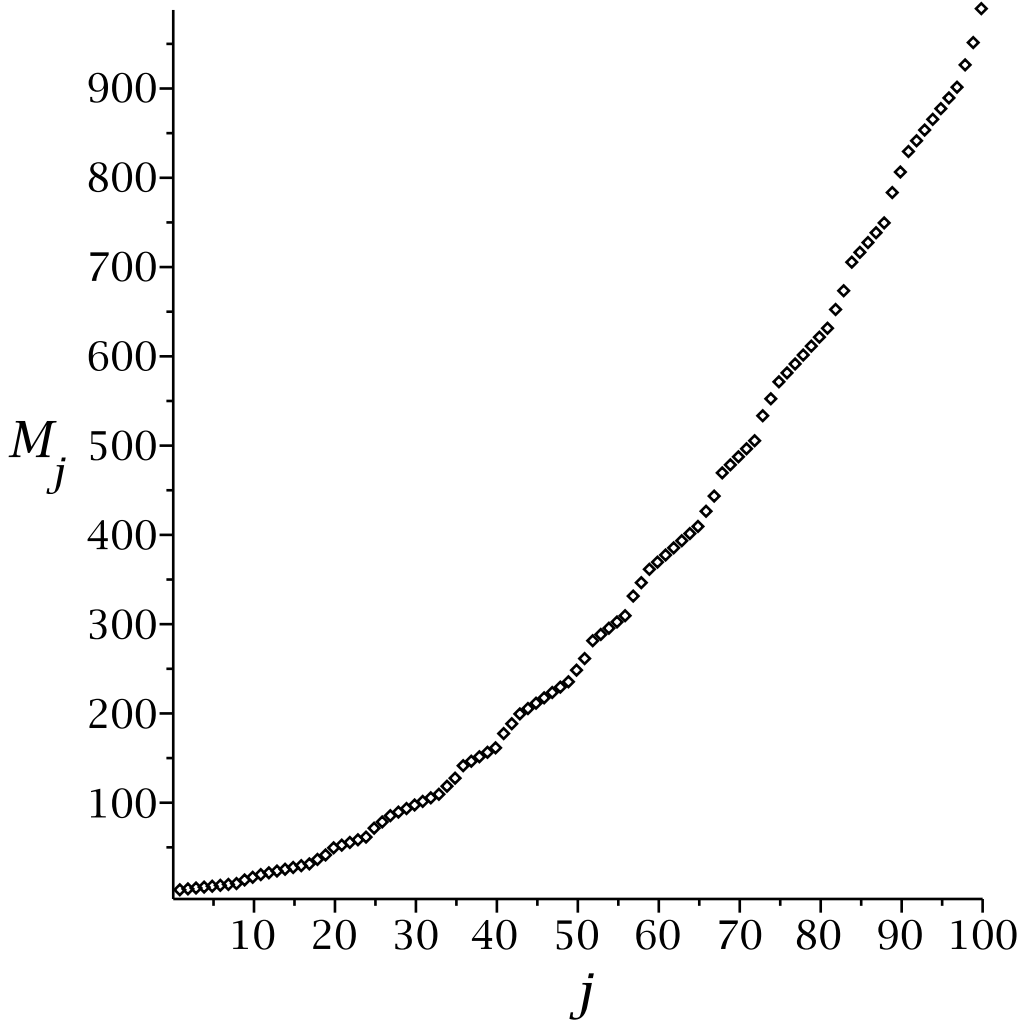}}
\end{center}
\else
\includegraphics[width=14cm]{racquets_M_j.eps}
\fi
\caption{\label{fig:M_j}
The sequence $\{M_j\}$ for the optimal $M_j$-racquets. }
\end{figure}

\begin{figure}                
\iftwocolumns
\begin{center}
\resizebox{\columnwidth}{!}{\includegraphics{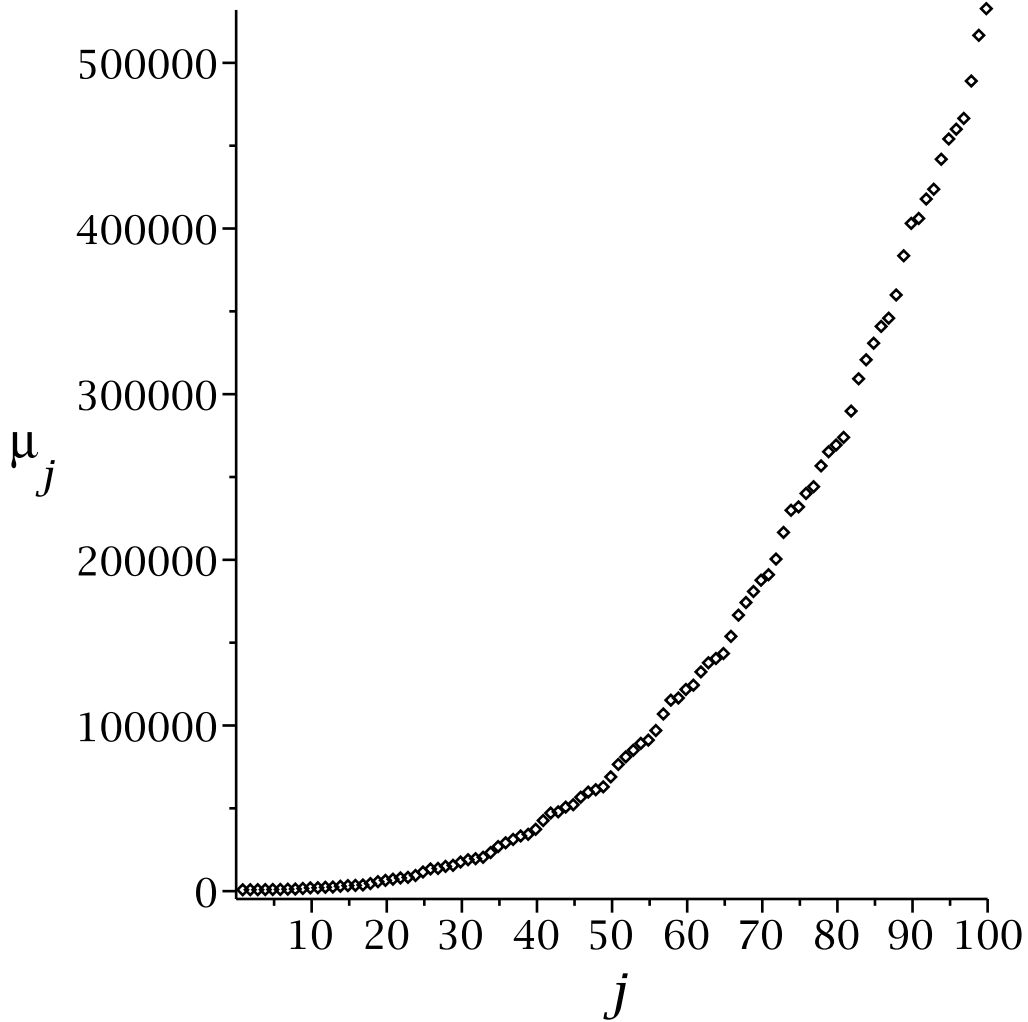}}
\end{center}
\else
\includegraphics[width=14cm]{racquets_mu_j.eps}
\fi
\caption{\label{fig:mu_j}
The sequence of critical lengths $\{\mu_j\}$ for the optimal $M_j$-racquets. }
\end{figure}

Next we compare the energy of the racquet-like shapes by using
Eq.~(\ref{eq:en_racq}) for $M$ and $Q$. We come to the equation for the critical value
of the effective length of the polymer $\mu \equiv \nu \lambda$ for the $M$- and $Q$-racquets
\begin{align*}
\mu_{MQ} = \frac{4\sqrt{2a} MQ}{M\alpha_{2Q}-Q\alpha_{2M}}\left[ \sqrt{M(\alpha_{M}-\alpha_{2M}+3M)}-\right. \\
\left.-\sqrt{Q(\alpha_Q-\alpha_{2Q}+3Q)}\right] .
\end{align*}
Similarly to the toroids we build a sequence of optimal racquets $\{M_j\}$ with critical lengths $\{\mu_i\}$. 
For a polymer of length $\mu$, the $M_j$-racquet is optimal among all racquet shapes
if $\mu_{j-1} < \mu < \mu_{j}$.
To start the sequence we set $M_0=0$, $\mu_0=0$, and $M_1=1$ corresponds to a single-covered two-headed racquet.
Then, assuming that we know the optimal sequence up to $j=J$ (i.e. we know
$M_j$, $j=0,\ldots, J$, and $\mu_{j}$, $j=0,\ldots, J-1$)
we find $\mu_J=\mu_{M_J Q^\star} =\min\limits_{Q>M_J} \mu_{M_J Q}$,
$M_{J+1}=\max\limits_{\mu_{M_I Q^\star}=\mu_J} Q^\star$.
As seen from Eq.~(\ref{eq:en_racq}), the energy of an $M$-racquet $u_r(\mu,M)$, for fixed $\nu$, is a linear function of $\mu$.
Moreover, it is an easy exercise to prove that the function
$M(\alpha_M-\alpha_{2M}+3M)$ monotonically increases with $M$ so that we have
$u_r(0,Q) > u_r(0,M)$ for $Q>M$. This means that if the graphs
$u_r(\mu,Q)$ and $u_r(\mu,M)$, $Q>M$, intersect each other at some positive $\mu_{MQ}$, then
$u_r(\mu,Q) < u_r(\mu,M)$ for $\mu > \mu_{MQ}$.
Therefore the $M_{J+1}$-racquet is optimal for $\mu > \mu_{J}$ in some right neighbourhood of $\mu_{J}$.
Now we iterate the procedure to compute $\mu_{J+1}$, $M_{J+2}$ and so on. 
The first members of the optimal racquet sequence are: 
$1,2,3,4,5,6,7,8,12,15,18,20,22,24,26,28,30,35\ldots$ (Fig.~\ref{fig:M_j}).
The critical lengths are shown in Fig.~\ref{fig:mu_j}.

The last comparison to make is between an $N$-covered circle and an $M$-racquet.
Equating energies taken from Eqs.~(\ref{eq:en_tor}) and (\ref{eq:en_racq})
leads to the quadratic equation
\begin{align}
&(3-\alpha_{2M}/(2M)) \mu^2 -2\sqrt{2aM(\alpha_M-\alpha_{2M}+3M)} \mu - \nonumber  \\
&-(3-\alpha_N/N)\lambda^2 +2\pi^2(N^2+c(N-1)^2) = 0.
\label{eq:torac}
\end{align}
Consider first the case $N=1$. The coefficient of the $\lambda^2$ term vanishes and we have a quadratic equation solely for $\mu$.
It has two positive roots but only the greater one is important:
$\mu_{1M} = \sqrt{2}(\sqrt{aM(\alpha_M-\alpha_{2M}+3M)}+\sqrt{aM(\alpha_M-\alpha_{2M}+3M)-\pi^2(3-\alpha_{2M}/(2M))})/(3-\alpha_{2M}/(2M))$.
Thus, the $M$-racquets win over simple circles if their length exceeds $\mu_{1M}$.
Clearly, we are most interested in $\mu_{11} = 2(\sqrt{2a}+\sqrt{2a-\pi^2}) \approx 22.4636$, 
because for $M \ge 2$ the simple circle is not competitive and the racquets should be compared with themselves and multicovered circles.
In the general case $N \ge 2$, Eq.~(\ref{eq:torac}) describes a hyperbola and it is convenient to resolve it for $\lambda$ as a function of $\mu$.
If the actual $\lambda$ exceeds this value, the energy of the $M$-racquet is higher than of the $N$-circle.

\begin{figure}                
\iftwocolumns
\begin{center}
\resizebox{\columnwidth}{!}{\includegraphics{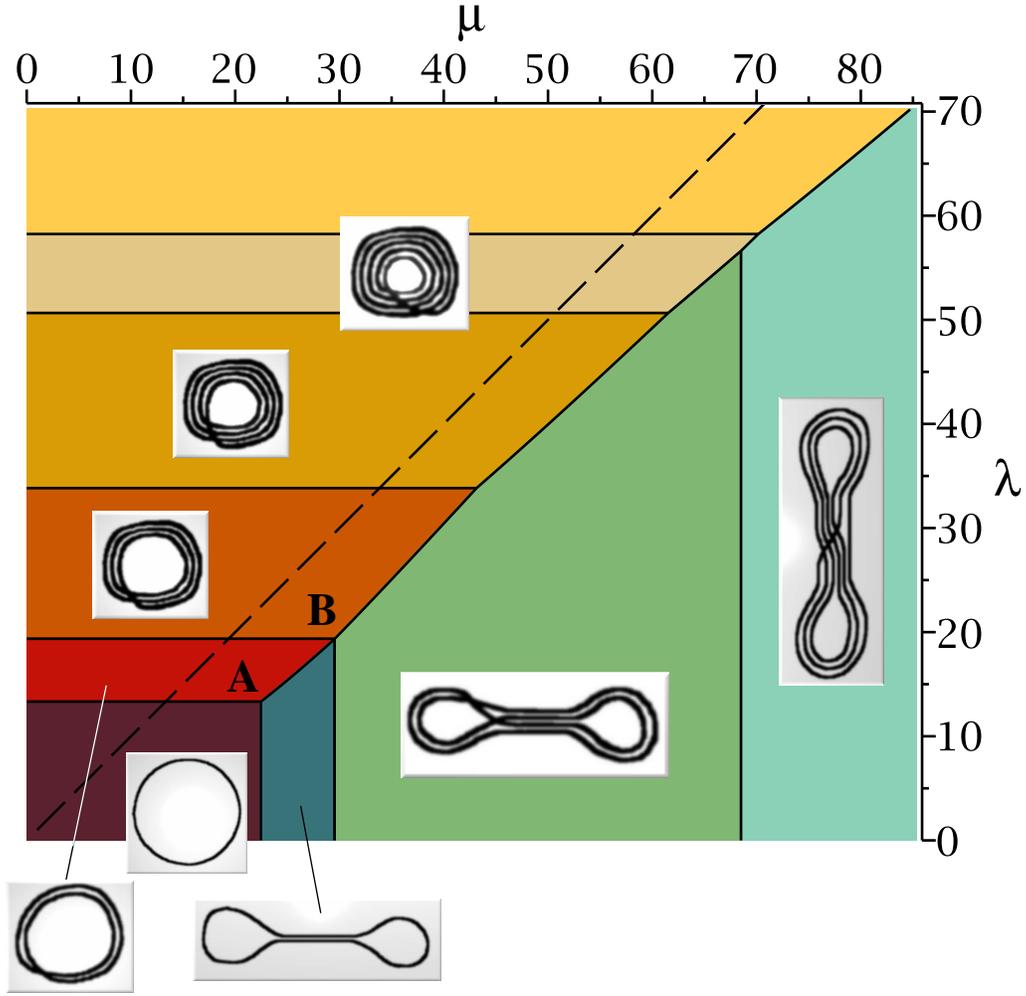}}
\end{center}
\else
\includegraphics[width=14cm]{diagrm2.eps}
\fi
\caption{\label{fig:diag} 
The diagram of state in the plane $(\mu, \lambda)$.
The domains where the conformations shown are ground states are painted in different colours. 
The bisector $\nu\equiv\lambda/\mu=1$ (dashed line) 
marks the case of invariant attraction forces independent of the monomer sequence.
The upper triangle $\nu<1$ corresponds to stronger attraction for the toroidal parallel arrangement of a polymer chain.
The lower triangle $\nu>1$ contains domains where racquets are ground states because their special geometry provides 
enhanced attraction.
The relative torsional stiffness is fixed $c=1.5$.}
\end{figure}

Our aim is to represent the results of energy comparison as a diagram of state in plane $(\mu, \lambda)$.
We begin with relatively short polymers (Fig.~\ref{fig:diag}).
A bisector divides the first quadrant of the parameter plane into two triangular domains.
The upper one, where $\lambda$ exceeds $\mu$, corresponds to $\nu^2 <1$, i.e. the stronger attraction for
toroidal arrangements of DNA. Clearly, the racquets have no chance to compete with toroids under such a condition unfavourable for them.
As the polymer is getting longer, the optimal condensate has a growing number of circle coverings.
The bisector itself corresponds to the equal strength of attraction potential irrespective of the DNA primary structure.
When the racquet-like shapes are benefited with $\nu > 1$, they appear on the surface.
In any case their length should exceed the minimum $\mu_{11}$.
With growth of the polymer chain, the optimal racquets acquire more dsDNA strands.
Similarly, toroids increase their number of coils as $\lambda$ becomes larger.
In the $(\mu, \lambda)$ plane the domain of the optimal $N_i$-toroid is bordered by
two parallel straight lines $\lambda=\lambda_{i-1}$ and $\lambda=\lambda_i$ with the neighbouring
$N_{i-1}$- and $N_{i+1}$-toroids. It also has a border with one or more racquet domains.
Each piece of this border is a part of a hyperbola (Eq.~(\ref{eq:torac})).
In a similar way, each domain of optimality of a given racquet type is bounded by two parallel lines $\mu=\mu_{j-1}$ and $\mu=\mu_j$
and one or more pieces of hyperbolae (Eq.~(\ref{eq:torac})).
Thus, the boundary between toroids and racquets is a piecewise curve made of hyperbolic intervals connecting triple points. 
The latter are either a toroid triple point where a straight line $\lambda=\lambda_i$
comes to, or a racquet triple point where two hyperbolae meet a straight line $\mu=\mu_j$.
The first triple point A ($\mu=\mu_{11}, \lambda=\lambda_1=2\pi\sqrt{3+c}$) is special, it corresponds
to the equality of energies of the single- and double-covered circles and the simplest racquet.

\begin{figure}                
\iftwocolumns
\begin{center}
\resizebox{\columnwidth}{!}{\includegraphics{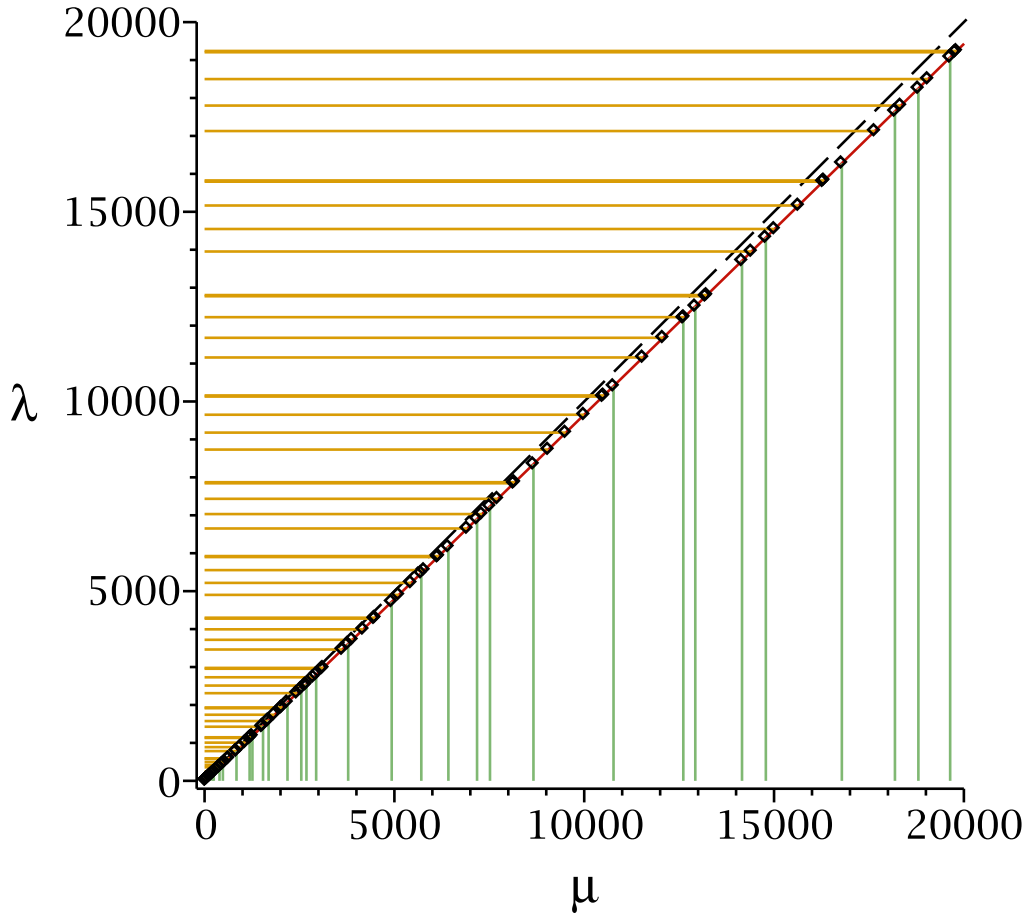}}
\end{center}
\else
\includegraphics[width=14cm]{diagram20000c1_5.eps}
\fi
\caption{\label{fig:diag20000c1_5} 
The diagram of state in the plane $(\mu, \lambda)$ for larger length ($c=1.5$).
Horizontal lines separate the toroids and vertical the racquets. Triple points are marked
on the toroid-racquet bondary curve. The dashed line is the bisector $\nu\equiv\lambda/\mu=1$.
}
\end{figure}

\begin{figure}                
\iftwocolumns
\begin{center}
\resizebox{\columnwidth}{!}{\includegraphics{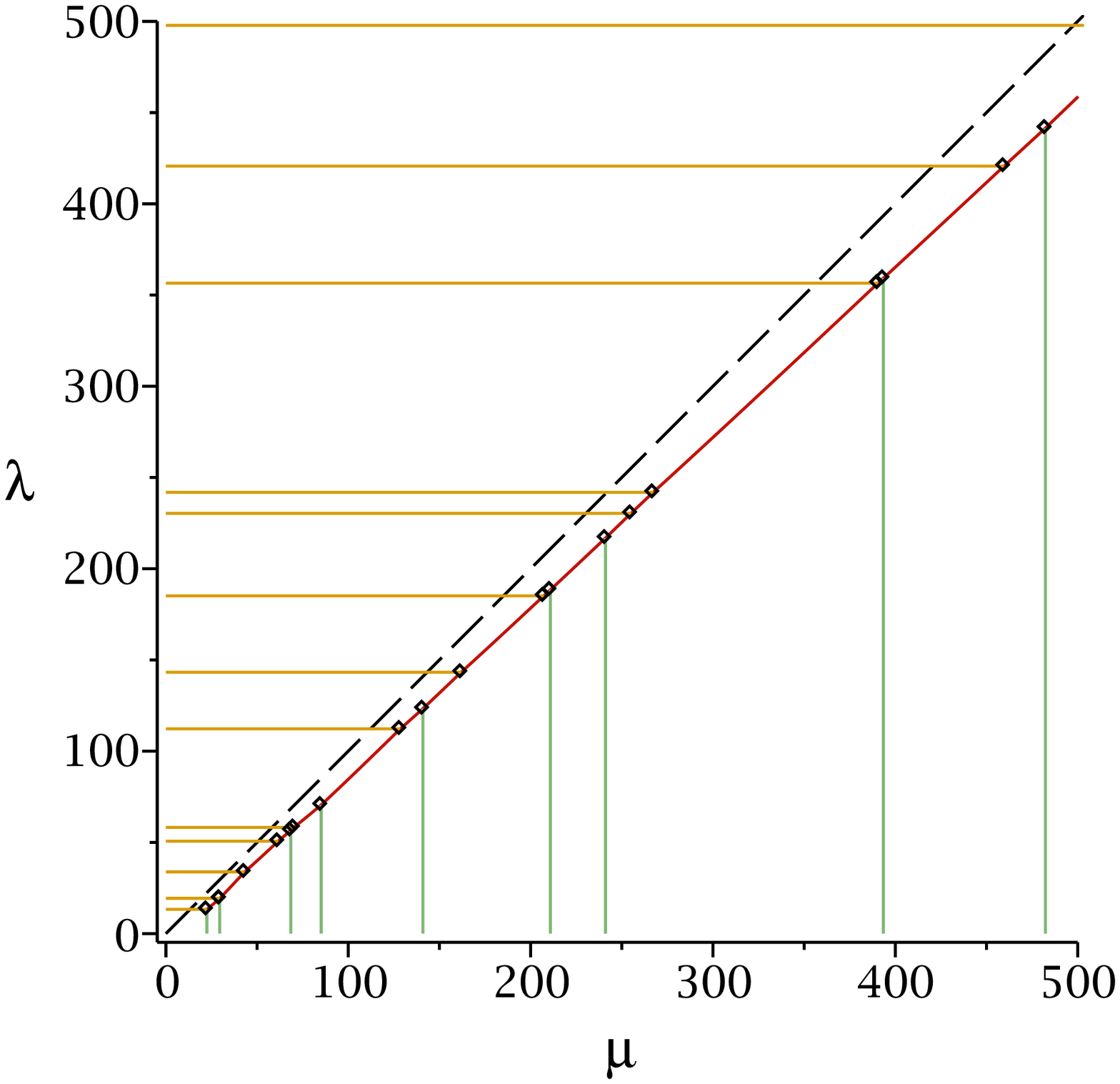}}
\end{center}
\else
\includegraphics[width=14cm]{diagram500c1_5.eps}
\fi
\caption{\label{fig:diag500c1_5} 
The diagram of state in the plane $(\mu, \lambda)$ ($c=1.5$) (a blowup of  Fig.~\ref{fig:diag20000c1_5}).
}
\end{figure}

As we have already computed the sequences $\{N_i\}$, $\{M_j\}$ and the corresponding critical lengths, we may construct the 
toroid-racquet boundary by applying the following algorithm. Suppose we know the boundary
that separates the $N_I$-toroids and $M_J$-racquets, that is a piece of the hyperbola that begins at 
a point $(\lambda_{I-1}, \mu^\star)$, $\mu^\star \ge \mu_{J-1}$, or $(\lambda^\star, \mu_{J-1})$, 
$\lambda^\star \ge \lambda_{I-1}$, depending on which coordinates are greater
(both points satisfy Eq.~(\ref{eq:torac}) for $N=N_{I}$, $M=M_{J}$). To fix the other end 
of our hyperbolic piece we have to choose from two candidates: (1) $(\lambda_{I}, \mu^{\star\star})$ or (2) $(\lambda^{\star\star}, \mu_{J})$
(again both points satisfy Eq.~(\ref{eq:torac}) for $N=N_I$, $M=M_J$). If $\lambda_{I} < \lambda^{\star\star}$ (and $\mu^{\star\star} <  \mu_{J}$),
 then the next triple point is toroidal (1) and the hyperbolic piece ends there, otherwise the end is the racquet triple point (2).
Then, to describe the next hyperbola piece by Eq.~(\ref{eq:torac}), we switch either to $N_{I+1}$ and $M_{J}$ in case (1) or to $N_{I}$ and $M_{J+1}$ in case (2).
Clearly, we can iterate this procedure to build the boundary curve up to arbitrary length (Figs.~\ref{fig:diag20000c1_5}, \ref{fig:diag500c1_5}).

The values of $\lambda_i$-s depend on the parameter $c$ so that for particular values of the latter some triple points
of different types may coincide. 
It looks on the diagram of state (Fig.~\ref{fig:diag}) like four boundary curves converge in one point B where
double- and triple-toroids and single- and double-racquets meet all together.
Actually there are a pair of triple points which are located indistinguishably close to each other 
because of the value of the parameter $c$ chosen. The quadruple point only exists for a slightly different value $c=1.4830$. 

\begin{figure}                
\iftwocolumns
\begin{center}
\resizebox{\columnwidth}{!}{\includegraphics{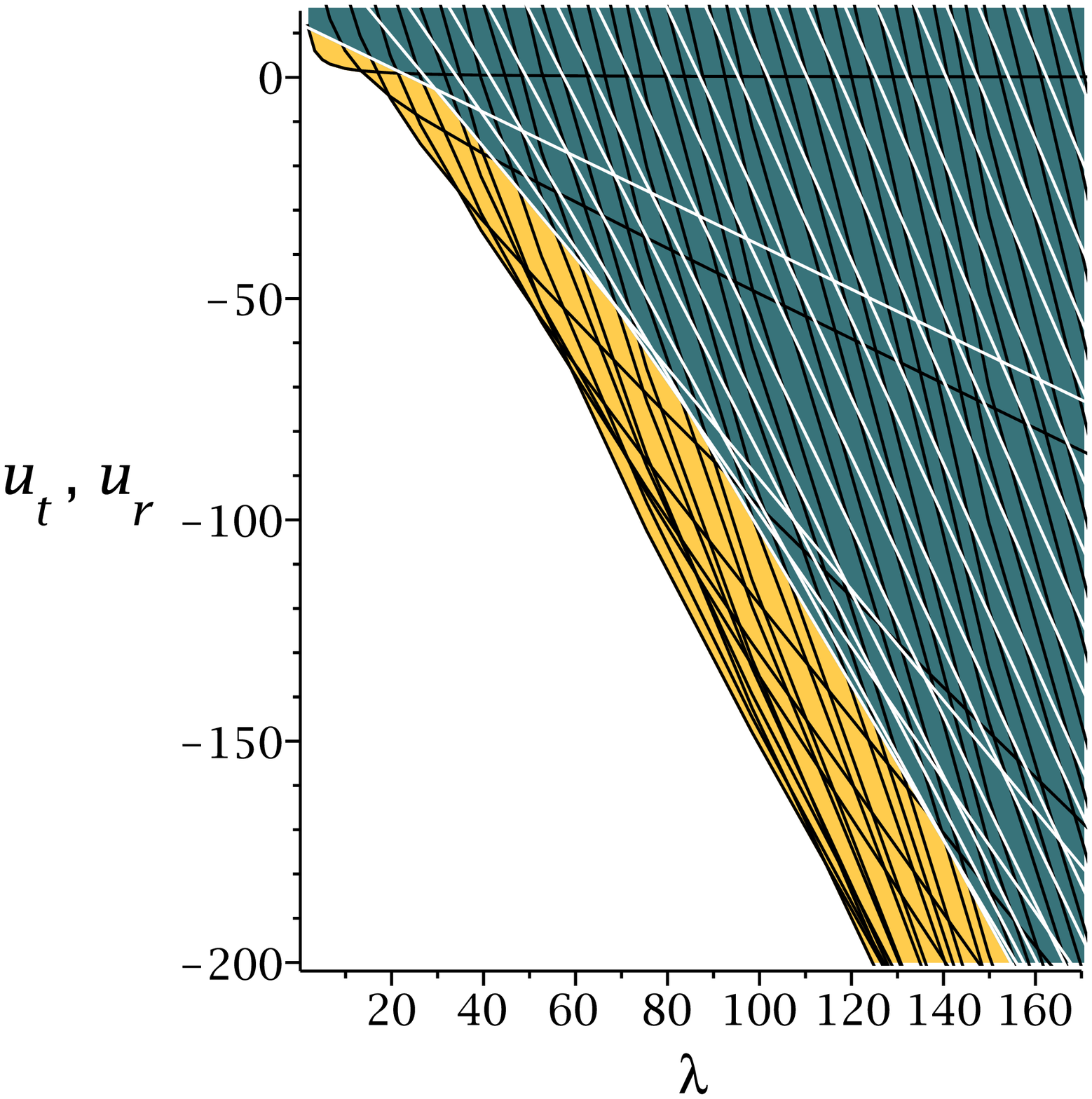}}
\end{center}
\else
\includegraphics[width=14cm]{energies_c1_5.eps}
\fi
\caption{\label{fig:energies_c1_5} 
Profiles of energies $u_t$ (black) and $u_r$ (white) (Eqs.~(\ref{eq:en_tor}) and (\ref{eq:en_racq}), resp.) for $c=1.5$, $\nu=1$.
The regions above the energy curves are painted yellow for the toroids and blue for the racquets.
}
\end{figure}

\begin{figure}                
\iftwocolumns
\begin{center}
\resizebox{\columnwidth}{!}{\includegraphics{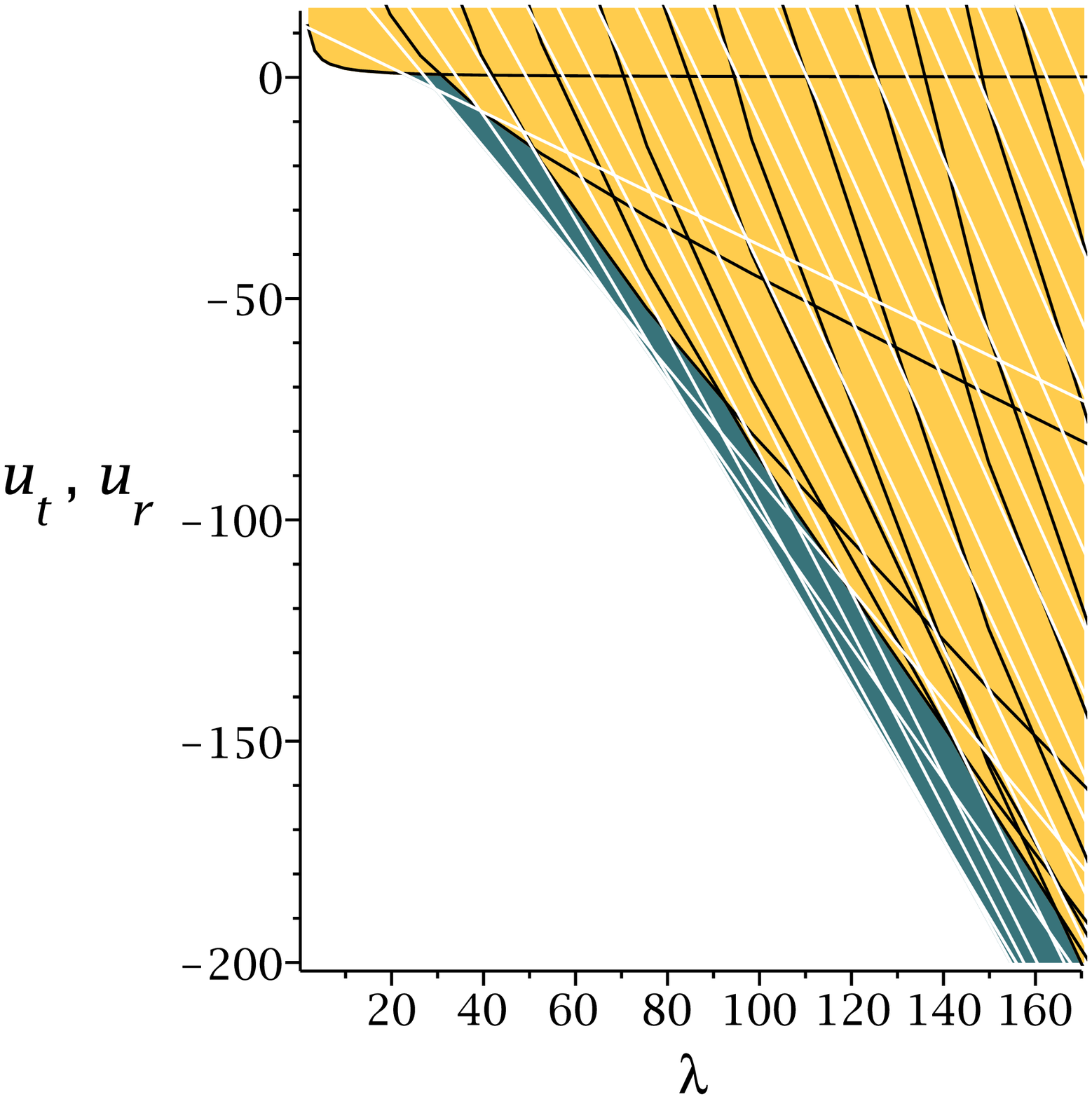}}
\end{center}
\else
\includegraphics[width=14cm]{energies_c20.eps}
\fi
\caption{\label{fig:energies_c20} 
Profiles of energies $u_t$ (black) and $u_r$ (white) (Eqs.~(\ref{eq:en_tor}) and (\ref{eq:en_racq}), resp.) for $c=20.0$, $\nu=1$.
The regions above the energy curves are painted yellow for the toroids and blue for the racquets.
}
\end{figure}

\begin{figure}                
\iftwocolumns
\begin{center}
\resizebox{\columnwidth}{!}{\includegraphics{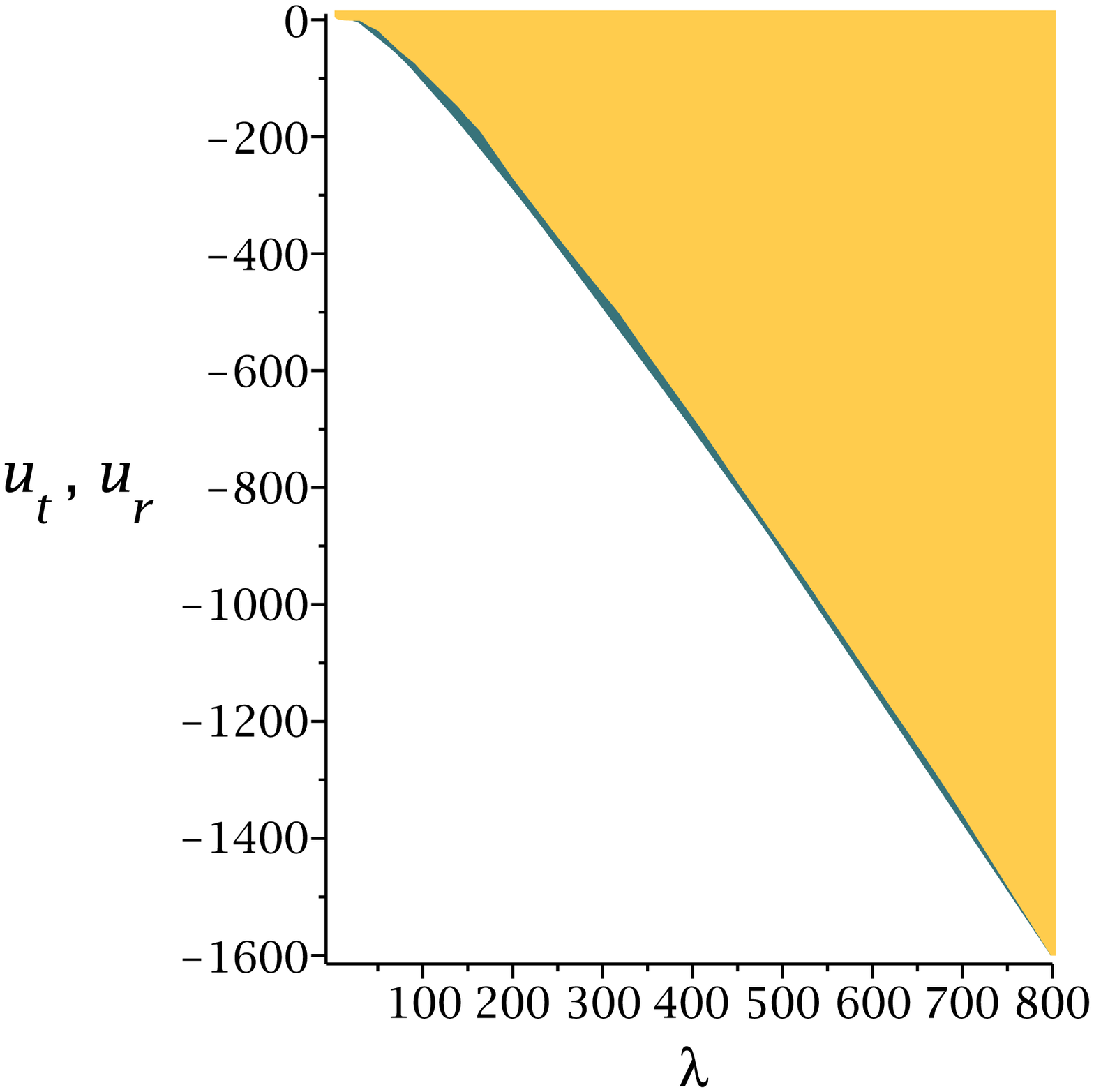}}
\end{center}
\else
\includegraphics[width=14cm]{energies_c20_800.eps}
\fi
\caption{\label{fig:energies_c20_l} 
Energies of the toroids $u_t$ (yellow) and the racquets $u_r$ (blue) (Eqs.~(\ref{eq:en_tor}) and (\ref{eq:en_racq}), resp.) for $c=20.0$, $\nu=1$,
for longer polymers.
}
\end{figure}

As we see in Figs.~\ref{fig:diag20000c1_5}, \ref{fig:diag500c1_5}, 
the boundary between the toroids and racquets remains under the bisector for given value of the
relative torsional stiffness $c$. We conclude then that if there is no difference between the interaction energy for
parallel and antiparallel alignment, then the racquet packing always loses to toroids.
Figure~\ref{fig:energies_c1_5} shows profiles of energies $u_t$ (black) and $u_r$ (white) (Eqs.~(\ref{eq:en_tor}) and (\ref{eq:en_racq}), resp.)
for $\nu=1$ (the bisector at the diagram of state in Figs.~\ref{fig:diag}--\ref{fig:diag500c1_5}), the regions above the energy curves
are painted yellow for toroids and blue for racquets to clearly demonstrate the difference. 
Note that though the torsional stiffness of DNA has not been measured with satisfactory precision~\cite{Kornyshev07},
the variation of $c$ seems to be within the range of $\sim0.5$ to $\sim 2.5$.
The boundary between the toroids and racquets remains below the bisector for such $c$.
However, if there exists another semi-flexible polymer with stronger torsional stiffness, then the racquets may have lower energy than toroids for some range of lengths. Indeed, as seen in Figs.~\ref{fig:energies_c20}, \ref{fig:energies_c20_l}, this occurs for $c=20$, but only
for a limited interval of lengths: $\lambda \lessapprox 800$.
Still, even for such a large $c$, the energy of racquets is not significantly lower than for toroids.

\section{Discussion}

The diagram of state is in agreement with the earlier conclusion that, in the zero-thickness model,
the racquet-like shapes can be only metastable, i.e.
they can only possess slightly greater energy than the toroidal-like conformations.
For open polymer shapes this was shown in Ref.~\cite{Schnurr02}.
Note that the 1-racquets (or hairpins) can turn out to be stable if the polymer (or a filament) has a significant thickness that cannot be ignored as for the chromatin fibre~\cite{Mergell04}.
However, as we have seen, the situation changes if the polymer interacts with itself differently depending on whether 
the sequences of the contacting base pairs are homologous or not. This sequence-dependent effect has been studied for DNA~\cite{Cherstvy04}.

The circular closedness constraint affects both the toroidal and racquet conformations.
Indeed, the racquets become less favourable because the polymer must make even number of end loops, one loop
more than in the minimal open racquet and each loop costs extra bending energy.
On the other hand, the non-zero writhe implies extra torsional energy for toroids.
The latter does not happen if the DNA is nicked and its linking number is not kept constant.
Setting zero torsional stiffness to model this case does not significantly change the diagram of state (Figs.~\ref{fig:diag}--\ref{fig:diag500c1_5}).

Another problem where a constrained semi-flexible polymer may form toroids is DNA under external tension in the presence of multivalent ions or polypeptides in the micromanipulation experiments~\cite{Battle09,Broek10,Marko12,Argudo12b}. It was found that, depending on the enviroment, toroids can compete with plectonemes.

We can estimate the condensation length for the B-form of DNA.
Its bending stiffness $B=k_B T l_p$, where $l_p \approx 50$ nm.
The DNA-DNA interaction
energy per length can be estimated as $\gamma \approx 0.1 k_B T\ \mbox{nm}^{-1}$~\cite{Kornyshev07}.
Thus, the DNA condensation length $L_c = \sqrt{B/\gamma} \approx 22$ nm at room temperature.
For $c=1.5$, the length that corresponds to the first triple point A is $L=\lambda_1 L_c \approx 300$ nm.
Thus the shortest DNA for which we expect the stable racquet is about $10^3$ bp long.
The parameter $\nu$ equals $\sim 1.7$ which is rather high, but smaller values are sufficient for the racquets made of longer DNA.

If the interacting parts happen to be homologous, then their interaction may be twice as strong as for nonhomologous molecules~\cite{Cherstvy04}.
This effect can be used to make the racquet shapes stable. To this aim,
the sequence may be chosen such that the homologous pieces come into contact in the particular racquet configuration but not in toroids.
Thus we would have $\gamma_r = \nu^2 \gamma = 2 \gamma$.
The effective racquet condensation length then shortens by factor $\nu$ and the normalised length $\mu$ increases by the same factor.
This effect is expected to be not so pronounced for multicovered conformations because 
the sequence correlations less affect the interaction energy in the tightly packed bundles~\cite{Cherstvy04}.

As we can see on the diagram of state, there is no chance for racquets if $\nu^2<1$ which describes 
the situation when the attraction is enhanced for parallel orientations of the contacting dsDNA intervals.
This effect could be easily achieved by building the closed DNA from homologous pieces joined so that their sequence orientation
is same.

On the other hand, if one of the two parts is reversed, then this would give the preference
to the antiparallel alignment like in the handle of the $1$-racquet and we have to look into the part of the diagram for $\nu^2>1$.

\section{Concluding Remarks}

The model considered in this paper assumes that the self-interacting sites tend to arrange themselves in a parallel bundle.
This is an idealisation that cannot be achieved in reality because of topological constraints~\cite{Starostin06}.
Nevertheless in many cases DNA forms highly-organised hexagonal bundles~\cite{Livolant96}.
This justifies the parallelism approximation.

The proposed tuning of the primary structure of dsDNA may lead to creation of intrinsically bent pieces, e.g. the so-called A-tracts. 
This is known to influence the size of toroidal condensates~\cite{Shen00} and
may also shift the balance between toroids and racquets because the competing conformations differ in curvature.
The latter is constant for toroids in the infinitely thin approximation but varies for racquets.

The dsDNA-dsDNA interaction potential is rather complex and it may have its minima for a relative orientation of the duplex axes at a non-zero skew angle
which depends on the distance between the molecules~\cite{Kornyshev07}.
This may cause formation of twisted bundles even for unconstrained polymers~\cite{Grason09}.
In the simplest case of a single pair of interacting molecules the complexity of the interaction forces may lead to emergence of a plectonemic structure which could cause
significant change of writhe. 
This effect can 
interfere with possible initial torsional strain of the closed DNA in its reference zero-writhe conformation~\cite{Ma94}.
To release this strain DNA may form plectonemes which could be facilitated or hindered by complex self-interaction forces.
Analysis of these effects is left for future work on elaboration of the model.

The present model uses a rough estimate of topological features of
two series of conformations focusing on the basic difference between two folding patterns.
The metastability of the racquet-like shapes and a possibility to finely control 
their conformational energy
may be exploited to construct a trigger mechanism which will be sensitive to small changes in the primary structure.

\vspace{5mm}
\begin{acknowledgments}
The author thanks G.H.M. van der Heijden, A. Korte, A.A. Kornyshev, D.J. Lee and R. Cortini for letting him bounce ideas off
their bright minds. 
Support of the UK's Engineering and Physical Sciences Research Council under grant no.
EP/H009736/1 is gratefully acknowledged. 
The colour palette in Fig.~\ref{fig:diag} is inspired by Wassily Kandinsky's ``Composition IX''
(1936), Mus\'ee National d'Art Moderne, Centre Georges Pompidou, Paris.
\end{acknowledgments}


%

\end{document}